\documentclass[%
 reprint,
superscriptaddress,
 amsmath,amssymb,
 aps,
]{revtex4-2}

\usepackage{amsmath}
\usepackage{physics}
\usepackage{ulem}
\usepackage[dvipsnames]{xcolor}
\usepackage{graphicx}
\usepackage{dcolumn}
\usepackage{bm}

\begin{document}

\preprint{APS/123-QED}

\title{Heat dissipation mechanisms in hybrid superconductor-semiconductor devices revealed by Joule spectroscopy}

\affiliation{Departamento de F\'{i}sica de la Materia Condensada, Universidad Aut\'{o}noma de Madrid, Madrid, Spain}
\affiliation{Departamento de F\'{i}sica Te\'{o}rica de la Materia Condensada, Universidad Aut\'{o}noma de Madrid, Madrid, Spain}
\affiliation{Condensed Matter Physics Center (IFIMAC), Universidad Aut\'{o}noma de Madrid, Madrid, Spain}
\affiliation{Instituto Nicol\'{a}s Cabrera, Universidad Aut\'{o}noma de Madrid, Madrid, Spain}
\affiliation{Center for Quantum Devices, Niels Bohr Institute, University of Copenhagen, Copenhagen, Denmark}

\author{A. Ibabe}
\affiliation{Departamento de F\'{i}sica de la Materia Condensada, Universidad Aut\'{o}noma de Madrid, Madrid, Spain}
\affiliation{Condensed Matter Physics Center (IFIMAC), Universidad Aut\'{o}noma de Madrid, Madrid, Spain}
\author{G. O. Steffensen}
\affiliation{Departamento de F\'{i}sica Te\'{o}rica de la Materia Condensada, Universidad Aut\'{o}noma de Madrid, Madrid, Spain}
\affiliation{Condensed Matter Physics Center (IFIMAC), Universidad Aut\'{o}noma de Madrid, Madrid, Spain}
\author{I. Casal}
\affiliation{Departamento de F\'{i}sica de la Materia Condensada, Universidad Aut\'{o}noma de Madrid, Madrid, Spain}
\affiliation{Condensed Matter Physics Center (IFIMAC), Universidad Aut\'{o}noma de Madrid, Madrid, Spain}
\author{M. G\'{o}mez}
\affiliation{Departamento de F\'{i}sica de la Materia Condensada, Universidad Aut\'{o}noma de Madrid, Madrid, Spain}
\affiliation{Condensed Matter Physics Center (IFIMAC), Universidad Aut\'{o}noma de Madrid, Madrid, Spain}
\author{T. Kanne}
\affiliation{Center for Quantum Devices, Niels Bohr Institute, University of Copenhagen, Copenhagen, Denmark}
\author{J. Nyg\r{a}rd}
\affiliation{Center for Quantum Devices, Niels Bohr Institute, University of Copenhagen, Copenhagen, Denmark}
\author{A. Levy Yeyati}
\affiliation{Departamento de F\'{i}sica Te\'{o}rica de la Materia Condensada, Universidad Aut\'{o}noma de Madrid, Madrid, Spain}
\affiliation{Condensed Matter Physics Center (IFIMAC), Universidad Aut\'{o}noma de Madrid, Madrid, Spain}
\affiliation{Instituto Nicol\'{a}s Cabrera, Universidad Aut\'{o}noma de Madrid, Madrid, Spain}
\author{E. J. H. Lee}
\email{eduardo.lee@uam.es}
\affiliation{Departamento de F\'{i}sica de la Materia Condensada, Universidad Aut\'{o}noma de Madrid, Madrid, Spain}
\affiliation{Condensed Matter Physics Center (IFIMAC), Universidad Aut\'{o}noma de Madrid, Madrid, Spain}
\affiliation{Instituto Nicol\'{a}s Cabrera, Universidad Aut\'{o}noma de Madrid, Madrid, Spain}

\date{\today}

%
%
%
%
%
%
%
%
%
%
%

      
\begin{abstract}

Understanding heating and cooling mechanisms in mesoscopic superconductor-semiconductor hybrid devices is crucial for their application in quantum technologies. Owing to the 
poor thermal conductivity of typical devices, heating effects 
can drive superconducting-to-normal phase transitions even at low applied bias, observed as sharp conductance dips through the loss of Andreev excess currents. Tracking such dips across magnetic field, cryostat temperature, and applied microwave power, which constitutes Joule spectroscopy, allows to uncover the underlying cooling bottlenecks in different 
parts of a device. By applying this technique, we analyze heat dissipation in devices based on full-shell InAs-Al nanowires and reveal that 
superconducting islands are strongly susceptible to heating as their cooling is limited by the rather inefficient electron-phonon coupling, as opposed to grounded superconductors that primarily cool by quasiparticle diffusion.
Our measurements indicate that powers as low as 50-150 pW are able to fully suprpress the superconductivity of an island. 
Finally, we show that applied microwaves lead to similar heating effects as DC signals, and 
explore the interplay of the microwave frequency and the effective electron-phonon relaxation time.

\end{abstract}

\maketitle

Hybrid superconductor-semiconductor nanostructures have attracted great attention in the past decade as a platform for the development of novel quantum devices \cite{Prada2020Oct, Aguado2020Dec}, with focus on both topological Majorana fermions \cite{Sau2010Jan, Lutchyn2010Aug, Oreg2010Oct}, and the study of conventional Andreev bound states \cite{Tosi2019Jan, Hays2021Jul, Wesdorp2023Sep, Pita-Vidal2023Aug}, among others. Additionally, of great interest are mesoscopic superconducting islands, comprised of 
tunnel coupled isolated regions with finite charging energy, which have been used to define charge qubits \cite{Bouchiat1998Jan, Makhlin2001May}, to detect topological transitions \cite{albrecht_exponential_2016, sherman_normal_2017, Valentini2022Dec}, and are promising for building artificial Kitaev chains \cite{Sau2012Jul, Dvir2023Feb, EstradaSaldana2022Apr}. Common for these applications is a sensitivity to heightened temperature and non-equilibrium distributions of quasiparticles, which could introduce both quasiparticle (QP) poisoning and qubit decoherence \cite{serniak_hot_2018, bargerbos_mitigation_2023, karzig_quasiparticle_2021, catelani_non-equilibrium_2019, dong_measurement_2022, le_calvez_joule_2019, de_cecco_interplay_2016, alegria_high-energy_2021}. The dynamics of such out-of-equilibrium QPs in superconductors, also referred to as "hot-electrons" \cite{wellstood_hot-electron_1994, pekola_single-electron_2013, zgirski_heat_2020, pothier_energy_1997} are also responsible for the observed micro-cooling in INSIN and SIN devices \cite{pannetier_quasiparticle_2009, pekola_trapping_2000, courtois_origin_2008, aref_andreev_2011, knowles_probing_2012} interesting for the pursuit of nanoscale thermodynamic elements beyond their detrimental effects in quantum computation. Despite great advances both in the fabrication and characterization of hybrid superconductor-semiconductor devices, the associated heating and relaxation mechanisms and their dependence on device geometry are still not well understood. Indeed, depending on the device and its environment, cooling can occur via different mechanisms, such as excited quasiparticles, limited at low temperatures by the superconducting gap, coupling to substrate phonons, which is strongly suppressed by the weak electron-phonon coupling at low temperatures, or emission of photons \cite{schmidt_photon-mediated_2004, meschke_single-mode_2006}, 
which requires an AC-generating coupling. Here, we study such effects for the case of hybrid nanowire devices with superconducting islands, which, as of their electric isolation, also display more limited cooling channels than in open hybrid systems, and therefore, could potentially be more susceptible to heating. 

While charge transport is rather straightforwardly measured through the application of voltage or current biases, 
the transport of thermodynamic heat is more subtle and requires both a well-controlled heat source and effective thermometers \cite{pekola_quantumheat_2021, pekola_thermometer_2018}. By applying a DC current across a device, Joule power is deposited at resistive elements and, at given power thresholds, can drive superconducting parts to the normal state, as the local temperature reaches the critical superconducting temperature, $T = T_c$. In devices with highly-transmitting junctions, such superconductor-to-normal transitions can be observed as a rapid drop of the superconducting excess current, measurable via lock-in techniques \cite{lee_above-gap_2010, tomi_joule_2021}. This, in total, yields both an electrically operated heat source and a single-shot thermometer at $T = T_c$,  which can be employed to investigate heating effects, a technique dubbed "Joule spectroscopy"  \cite{ibabe_joule_2023}. Furthermore, if $T_c$ is tunable, for example via an external magnetic field, this single-shot measurement can be extended to a wider range of temperatures and powers.   

        \begin{figure*}[t]
        \includegraphics{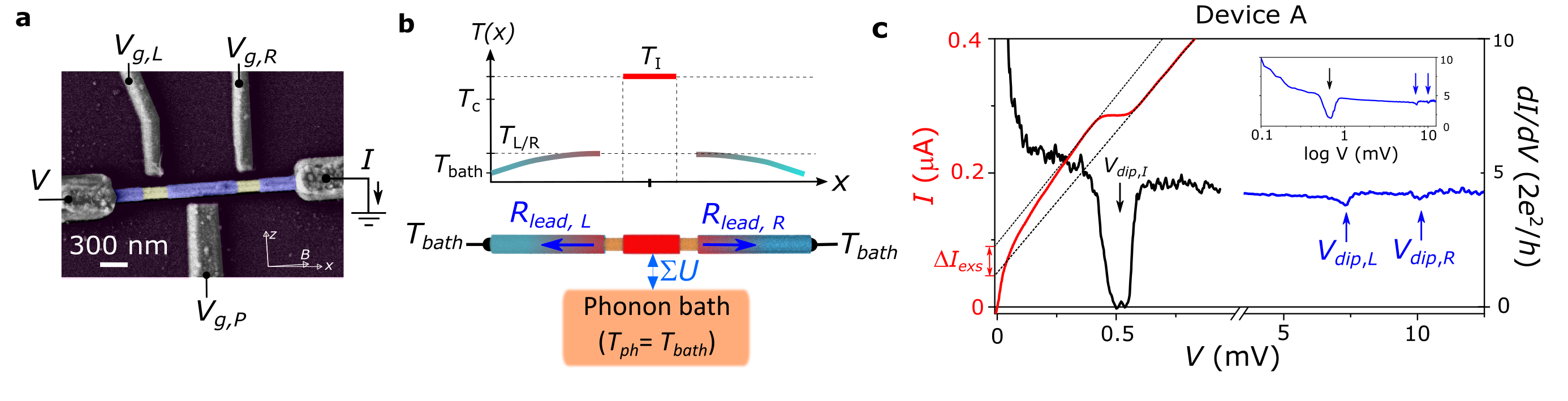}
        \caption{\label{Fig1}  $\mid$
        {\textbf{
        Joule heating in mesoscopic superconducting islands.}} \textbf{a},  Electron micrograph (false color) of a device  similar to device A. A superconducting island of length $L \approx 0.85$ $\mu$m and two superconducting leads (purple) are defined by wet etching two segments of the Al shell of an InAs-Al nanowire. 
        Side gates $V_{g, L}$ $V_{g, P}$, and $V_{g, R}$ are fixed at 5 V for the measurements.
        An external magnetic field, $B$, is applied with a small angle with respect to the NW axis, $x$. \textbf{b}, Schematics of the temperature distribution along the device due to Joule heating. $T_{bath}$ is taken as the cryostat temperature, whereas $T_{j}$  (where $j$ = $I, L, R$) are the temperatures at the island and on the left and right leads. Our thermal model (lower panel) indicates the main cooling mechanism for each superconductor: quasiparticle diffusion for the leads (governed by the parameter $R_{lead,i}$), and electron-phonon coupling $\propto \Sigma U$ for the islands. \textbf{c}, $I(V)$ (red line) and $dI/dV(V)$ (black and blue lines) curves taken at $B = 0$. Three $dI/dV$ dips are observed with increasing $V$, as indicated by the arrows. The dips signal superconductor-to-normal metal transitions of the island at  $V_{dip, I} \approx 0.5$ mV and  the leads, $V_{dip,L} \approx$ 7.5 mV and $V_{dip,R} \approx$ 10 mV. The $I(V)$ curve shows a suppression of the excess current, $\Delta I_{exs}$, when the island turns normal. 
        Inset: $dI/dV(V)$ measurement plotted against voltage in log scale. The black (blue) $dI/dV$ curves were taken with $dV =$ 5$\mu$V (200 $\mu$V) to better resolve the low- (high-) bias dips (see Methods for more detail).  
        }
        \end{figure*}

        \begin{figure}[t]
        \includegraphics{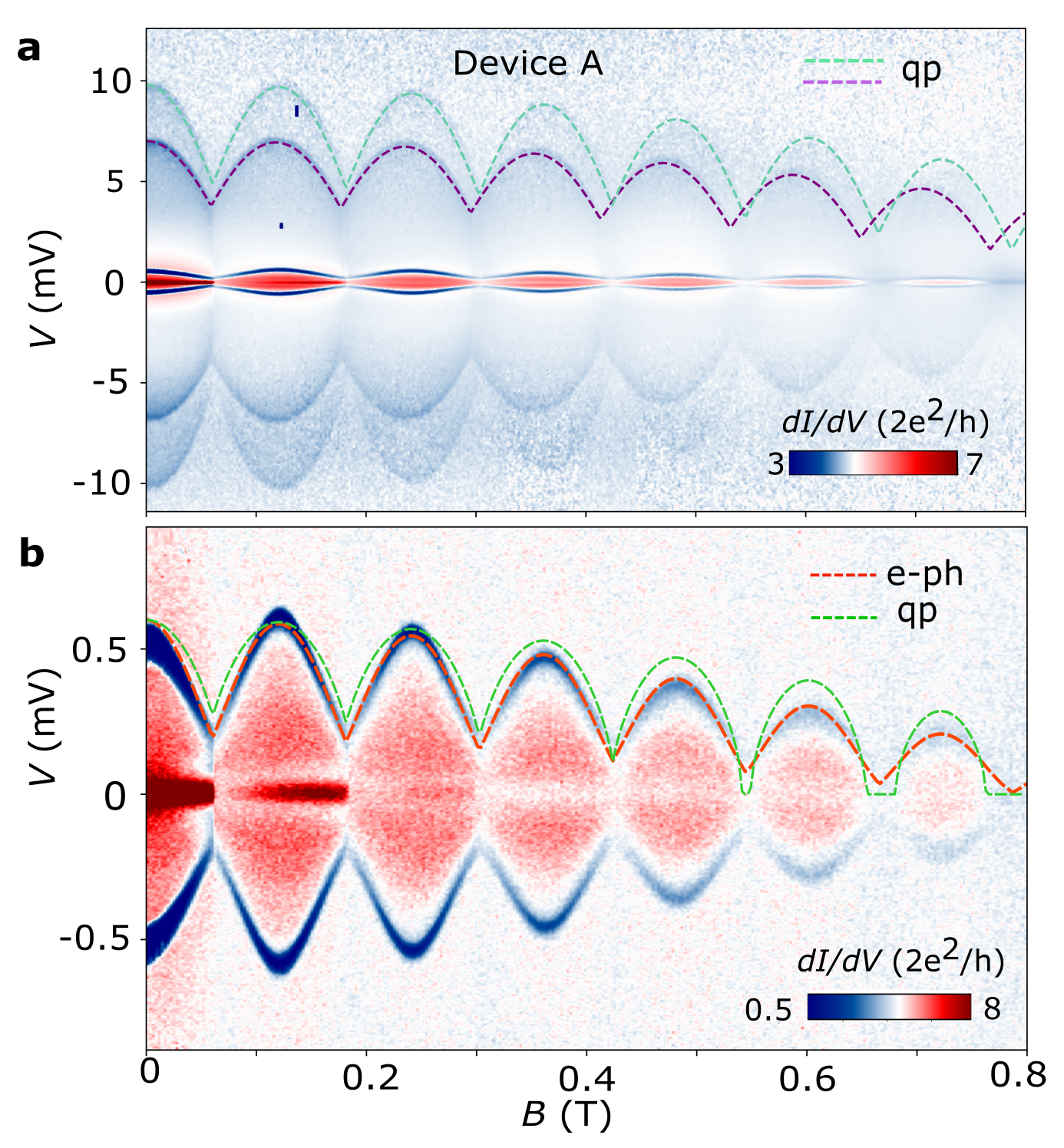}
        \caption{\label{Fig2}  $\mid$
        {\textbf{Heat dissipation mechanisms revealed by Joule spectroscopy.}} \textbf{a}, $dI/dV(V)$ as a function of magnetic field for device A, taken at a cryostat temperature of 20 mK. Owing to the Little-Parks effect, the positions of the three $dI/dV$ dips oscillate with applied $B$. The dependence of the higher bias dips, $V_{dip, L/R}(B)$, which relate to the superconductor-to-normal transitions of the two full-shell InAs-Al leads, is directly proportional to $T_c(B)$ as underscored by the fits to the Abrikosov-Gork'ov theory (green and purple dashed lines, labeled as qp, referring to quasiparticle diffusion). By contrast, the magnetic field dependence of the lower bias dip, ($V_{dip, I}$), in \textbf{b}, is clearly distinct (the green dashed line is an attempt to fit the experimental data to $V_{dip, I}(B) \propto T_c(B)$). A good agreement with the experimental data is obtained for $V_{dip, I}(B) \propto \sqrt{T_c^{6}(B)}$ (red dashed line, labeled as e-ph, referring to electron-phonon coupling).}
        \end{figure}

In this work, we investigate the relevant heat dissipation mechanisms in hybrid superconductor-semiconductor nanowire devices and reveal that 
grounded and floating superconductors display distinct heat bottlenecks. 
We focus on devices based on full-shell InAs-Al nanowires incorporating a mesoscopic superconducting island and present data corresponding to two devices in the main text. The first one, referred to as device A, was fabricated by wet etching two segments of approximately $200$ nm of the Al shell, thus defining an island of length $L \approx 0.85$ $\mu$m and two superconducting full-shell InAs-Al leads. Fig.~\ref{Fig1}a displays a scanning electron micrograph of a device lithographically similar to device A (see Methods for a detailed description of the fabrication and of the different devices).
Side gates $V_{g, L}$ and $V_{g, R}$ tune independently the transparency of each nanowire junction, 
and the plunger gate, $V_{g, P}$, was fixed at 5 V for all measurements. As Joule spectroscopy relies on detecting changes in the excess current, we have carried out measurements in the open regime to maximize $I_{exs}$. Specifically, we fix $V_{g, L}$ = $V_{g, R}$ = 5 V, obtaining a normal conductance $G_N \approx 4.5 \times 2e^2/h$ and $I_{exs} \approx$ 95 nA. We note that, in this regime, charging effects are also suppressed and for this reason they are neglected in the remaining of our analysis. Finally, the application of an external magnetic field, $B$, which has a small angle of $\approx 3^{\circ}$ with respect to the nanowire in device A, provides a powerful knob in our study. Indeed, the modulation of the superconducting critical temperatures, $T_c$, of the island and of the leads by the Little-Parks effect \cite{LittleParks1962, Vaitiekenas2020} will allow us to distinguish between different mechanisms for heat dissipation in our device.


Let us start by discussing the impact of Joule heating on the temperature of hybrid superconductor-semiconductor devices. At the core of this problem lies the thermal balance between the amount of heat dissipated by the Joule effect and the existing cooling power in the device. Importantly, both heating and cooling powers may be distributed rather unevenly, as is the case for devices with superconducting islands. 
It is therefore useful to analyze the thermal balance at each of the superconductors separately:
\begin{equation}
\label{heatbal}
 P_{h, j} =  P_{c, j}(T_j, T_{bath}),
\end{equation}
where $P_{h, j}$ and $P_{c, j}(T_j, T_{bath})$, respectively, refer to the Joule power deposited on, and the cooling power available at superconductor $j$ (with $j = I, L, R$ corresponding to the island, left lead and right lead). 
Note that the latter depends both on the electronic temperature of the superconductor, $T_j$, and the bath temperature, $T_{bath}$, which we take to be equal to the cryostat temperature.

The heating considered here arises from the relaxation of hot quasiparticles generated in the superconductors by the flow of dissipative electrical currents. For the sake of simplicity, we first discuss this effect for the case of a highly-transmitting $S-S$ junction. For $V > (\Delta_L + \Delta_R)/e$, transport is well described by $I = V/R + I_{exs,L} + I_{exs,R}$, where $\Delta_L$ and $\Delta_R$ are the superconducting gaps of the left and right leads, and $R$ is the normal resistance. $I_{exs, L}$ ($I_{exs, R}$) refers to the excess current stemming from Andreev reflections on the left (right) lead. As a result, the dissipated Joule power will have contributions from both Ohmic and excess current terms above. Interestingly, while the former is distributed evenly between left and right leads, the excess current contributions are not. This owes to the fact that the Andreev reflections underlying $I_{exs}$ generate quasiparticles only on the opposite side of the junction. Hence, the Joule power deposited on the left (right) lead is given by $P_{h, L} = V^2/2R + I_{exs, R}V$ ($L\leftrightarrow R$). By extending the same logic to the geometry of device A (see Supplemental Material for more details), we get:
\begin{align}
I &= \frac{V}{R}+\frac{1}{2}\left(2I_{exc,I}+I_{exc,L}+I_{exc,R}\right), \label{eq:TotCurr} \\
P_{h, I} &= \frac{1}{2}\frac{V^2}{R} + \frac{V}{2}\left(I_{exs,L}+I_{exs,R}\right), \label{eq:IslandHeat} \\
P_{h, L/R} &= \frac{1}{4}\frac{V^2}{R} + \frac{V}{2}I_{exs,I}, \label{eq:LeadPow}
\end{align}

where we assumed that the resistances of the two nanowire junctions are equal, $R_L = R_R = R/2$, leading to approximately equal voltage drops across each junction, $V/2$. Note that the relations above satisfy the detailed balance, $IV = P_{h, I} + P_{h, L} + P_{h, R}$, and demonstrate that the Joule power deposited on the superconductors can be different.

At thermal equilibrium, the power deposited on superconductor $j$ has to match $P_{c, j}(T_j, T_{bath})$, the strength of which increases with $T_j$. This means that if at a given instant $P_{h, j} > P_{c, j}(T_j, T_{bath})$, then $T_j$ rises to restore the balance in Eq.~(\ref{heatbal}). In this way, upon increasing the Joule power, an inhomogeneous temperature distribution builds up in the device (Fig.~\ref{Fig1}b), with hotter zones forming in regions with less efficient heat dissipation and/or a higher Joule power deposition. Clearly, for high enough Joule powers, the temperature can reach the superconducting critical temperature at a given superconductor, i.e., $T_j = T_{c,j}$, driving its transition to the normal state. This occurs at a critical voltage, $V_{dip,j}$, and leads to the full suppression of Andreev processes and  $I_{exs,j} = 0$. Such a drop in current results in a sharp dip in differential conductance, $dI/dV$, at $V_{dip,j}$, which is measured in the experiment. We detect three such $dI/dV$ dips for device A (Fig.~\ref{Fig1}c): one at relatively low bias ($V_{dips, I} \approx 0.5$ mV), which we ascribe to the superconducting island as it is more thermally isolated, and two appearing at higher bias ($V_{dip,L} \approx$ 7.5 mV and $V_{dip,R} \approx$ 10 mV), attributed to the left and right leads. A partial suppression of the excess current, $\Delta I_{exs}$, which correlates with the low-bias dip, is clearly identified in the $I(V)$ curve in red, with $I_{exs}$ finally going to zero for $V \gtrsim$ 10 mV (not shown). 

By applying Eqs. (3) and (4), we estimate the Joule powers required for driving the island and the leads of device A to the normal state, $P_{dip,I} \approx$ 60 pW and $P_{dip, L} \approx$ 5 nW and $P_{dip, R} \approx$ 9 nW, respectively. Such a difference of around two orders of magnitude already suggests that the underlying heat dissipation mechanisms must be different. Following previous works \cite{tomi_joule_2021, ibabe_joule_2023}, we ascribe quasiparticle diffusion as the dominant mechanism for the superconducting leads. Solving the corresponding heat diffusion equation yields the cooling power at $T_c$ for the leads,
\begin{equation}
P_{dip,j} = P_{c,j}(T_{c,j}) = \Lambda \frac{k_B^2 T_{c,j}^2}{e^2 R_{lead,j}}, 
\end{equation}
with $R_{lead,j}$ being the normal resistance of the leads, and $\Lambda = 2.112$ \cite{ibabe_joule_2023}. By contrast, owing to the presence of the nanowire junctions, the superconducting island can not cool down efficiently via quasiparticles. Instead, we assume that the heat dissipation predominantly occurs by electron-phonon coupling,
\begin{equation}
P_{dip,I} = P_{c,I}(T_{c,I}, T_{bath}) = \Sigma U \left(T_{c,I}^n - T_{bath}^n\right) \label{eq:PhCooling},
\end{equation}
where $\Sigma$ is a material-dependent parameter, $U$ is the volume of the island, and $n$ is expected to range from 4 to 6, according to theory \cite{reizer_electron-phonon_1989, wellstood_hot-electron_1994, lemzyakov_electron-phonon_2022, oneil_measurement_2012, subero_bolometric_2022, karvonen_observation_2005, underwood_insensitivity_2011}. Note that, due to their limited impact, other possible heating and cooling terms are neglected at this moment. 


        \begin{figure*}[t]
        \includegraphics{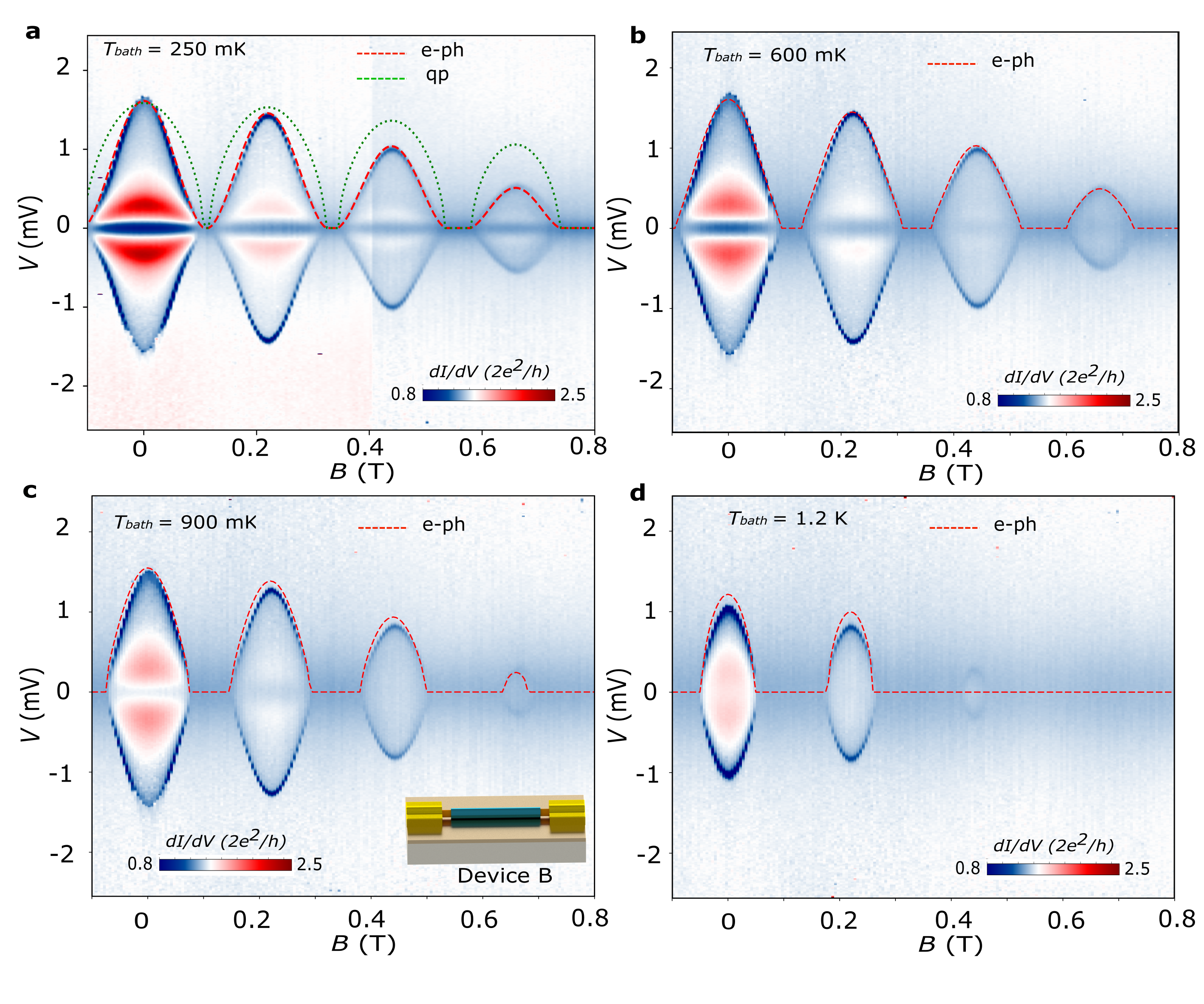} \caption{\label{Fig3} $\mid$ \textbf{Electron-phonon coupling as the dominant heat dissipation mechanism in superconducting islands.} 
        $dI/dV(V)$ as a function of the external magnetic field for device B at 
        different cryostat temperatures: \textbf{a}, 250 mK, \textbf{b}, 600 mK, \textbf{c}, 900 mK, and \textbf{d}, 1.2 K. In this dataset, a single $dI/dV$ dip related to the island is observed, as the leads in this device are normal (Cr/Au, see inset in panel \textbf{b}). 
        For the measurement taken at the lowest temperature (panel \textbf{a}), we plot the best fits to $V_{dip, I} \propto T_{c}$ (green dashed line, labeled as qp) and $V_{dip, I} \propto \sqrt{(T_{c}^6-T_{bath}^6})$ (red dashed line, labeled as e-ph), using AG theory. $T_{bath}$ is taken as the cryostat temperature. For the other panels, we plot $V_{dip, I}(B)$ calculated using the same fit parameters extracted from \textbf{a} and the corresponding $T_{bath}$. 
        } 
        \end{figure*}

To test our assumptions regarding the dominant heat dissipation mechanisms for the superconducting island and leads, we study the dependence of the three $dI/dV$ dips as a function of a magnetic field applied nearly parallel to the nanowire axis, as shown in Fig.~\ref{Fig2}. Owing to the Little-Parks effect, the magnetic field gives rise to $T_{c,j}(B)$ oscillations which translate into characteristic $V_{dip,j}(B)$ dependences for the distinct dissipation mechanisms.
For example, quasiparticle diffusion yields $V_{dip,j}$ directly proportional to $T_{c,j}$, as concluded from Eqs. (4) and (5). We thus fit the high-bias dips in Fig.~\ref{Fig2}a to $V_{dip,L/R}(B) \propto T_{c,L/R}(B)$ using Abrikosov-Gorkov (AG) theory \cite{AbrikosovGorkov1961, ibabe_joule_2023}. The excellent observed agreement reinforces our interpretation of the dominant cooling mechanism for the leads at $T_c$. Note that the fact that the two high-bias dips are not identical is due to asymmetries in the device, e.g., in $R_{lead,j}$. The $B$-field behavior of the low-bias dip, on the other hand, does not follow the same dependence, as evidenced by the green dashed line in Fig.~\ref{Fig2}b.  Instead, we fit the experimental data to $V_{dip,I}(B) \propto \sqrt{T_{c,I}(B)^{n}-T_{bath}^{n}}$, which is obtained from Eqs. (3) and (6) by neglecting the excess current terms in $P_{h,I}$ (which correspond to approximate $25 \%$ of total $P_{h,I}$ in device A, and may explain the lower $V_{dip,I}$ observed in lobe $0$ compared to lobe $1$ in Fig.~\ref{Fig2}b, see supplemental material). A good agreement is achieved for $n = 6$ (red dashed line), with $T_{c,I}(B)$ obtained from AG theory and $T_{bath} = 0$, thus supporting our assumption of electron-phonon coupling as the main cooling mechanism in mesoscopic islands.


To further scrutinize heat dissipation in the islands, we have studied a second device, whose geometry incorporates two main modifications. First, for this device B, we employ normal metal leads, which simplifies our analysis of the superconductor-to-normal transition of the island as the excess current terms in Eq.~(\ref{eq:IslandHeat}) disappear. The device is fabricated by etching the Al shell at the ends of the nanowire and evaporating the Cr/Au leads, thus defining a $L\approx 1.4$ $\mu$m-long island (see schematics in the inset of Fig.~\ref{Fig3}b). Second, the device is fabricated with a thinner InAs-Al nanowire which enters the destructive regime of the Little-Parks effect. This is useful as the differences between the quasiparticle diffusion and electron-phonon coupling dependences for $V_{dip,I}(B)$ grow with shrinking $T_{c,I}$ as already observed for device A. To maximize the excess current, measurements were taken again in the open regime (where $V_{g}$ = 10 V is the voltage applied to the global back gate, $G_N \approx 1.6 \times 2e^2/h$, $I_{exs} \approx$ 25 nA). As device B only contains one superconducting element, a single $dI/dV$ dip is observed, consistent with the full suppression of $I_{exs}$ when the island turns normal. Fig.~\ref{Fig3} displays $V_{dip,I}(B)$ measurements taken at different cryostat temperatures, $T_{bath} =$ 250, 600, 900 and 1200 mK. Note that here $V_{dip,I} \approx 1.5$ mV and $P_{dip,I} \approx$ 150 pW at $B=0$ and $T_{bath}= 250$ mK. 
We fit the experimental data in Fig.~\ref{Fig3}a to $V_{dip,I}(B) \propto T_{c,I}(B)$ (green dashed line), and to $V_{dip,I}(B) =  \beta \sqrt{T_{c,I}(B)^{6}-T_{bath}^{6}}$ (red dashed line), where $\beta = \sqrt{2R \Sigma U}$ is a fitting parameter, $T_{bath}$ = 250 mK, and $T_{c,I}(B)$ is obtained by AG theory. An excellent agreement is obtained with the latter, from which we estimate $\Sigma \approx 4.3 \times 10^{9}$ W/m$^3$K$^6$ by taking $R$ = 8.1 k$\Omega$, and $U \approx 4 \times 10^{-21}$ m$^3$ as the volume of the Al shell in the island. The estimated $\Sigma$ is in good agreement with previous works \cite{karvonen_observation_2007, subero_bolometric_2022}. In the other panels, we plot $V_{dip,I}(B)$ calculated using $\beta$ extracted above, and the corresponding cryostat temperature as $T_{bath}$ (dashed lines). 
The very good agreement for all measurements strongly supports our interpretation of electron-phonon coupling as the main cooling mechanism in islands at $T_c$. Note that slight discrepancies in the measured and calculated $V_{dip,I}$  appear for higher $T_{bath}$, most clearly seen for the measurement taken at 1.2 K. This could be related, e.g., to the impact of other weaker cooling/heating mechanisms neglected in our analysis, and/or to temperature-dependent electron-phonon coupling parameters, e.g., $\Sigma$ and $n$ \cite{underwood_insensitivity_2011}.


        \begin{figure}[t]
        \includegraphics{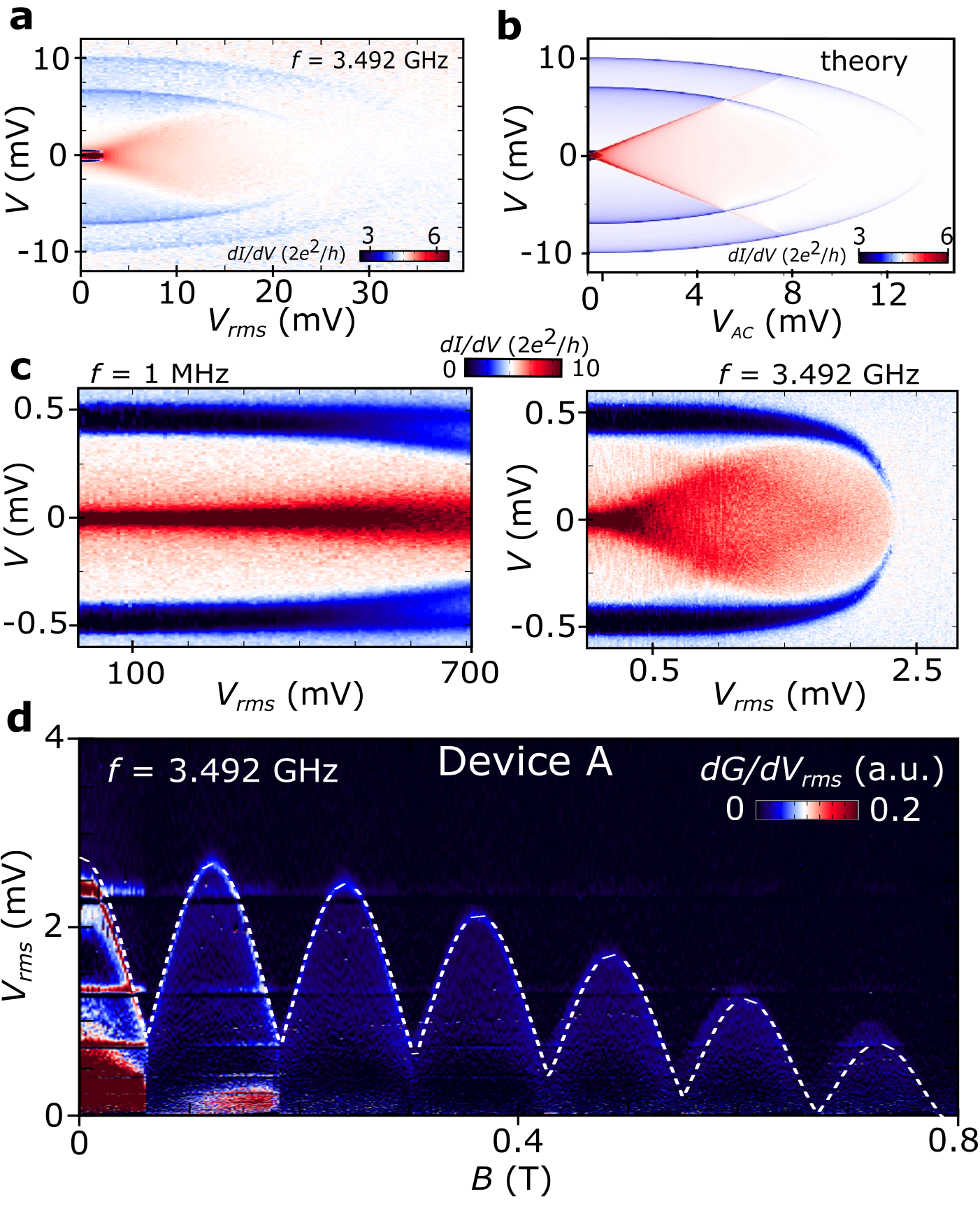} 
        \caption{\label{Fig4} $\mid$
        \textbf{Heating effects with high-frequency AC signals.} \textbf{a}, $dI/dV(V)$ of device A as a function of the nominal rms amplitude, $V_{rms}$, of a $f = $ 3.492 GHz AC signal applied to a nearby antenna (see Methods for more detail).  
        \textbf{b}, Theoretical simulation of $dI/dV(V)$ as a function of the amplitude of an AC voltage signal, $V_{AC}$, superimposed to the DC bias, $V$. \textbf{c}, Impact of the AC signal frequency on the heating of the island. $dI/dV(V)$ as a function of $V_{rms}$ for $f = $ 1 MHz (left panel) and 3.492 GHz (right panel). For the lower frequency, a splitting of the low-bias $dI/dV$ dip is observed. \textbf{d}, $B$-field dependence of the microwave-induced superconductor-to-normal transition of the island at zero DC voltage. To improve the visibility of the transition, we plot the numerical derivative of the zero-bias conductance, $G = dI/dV(V=0)$, with respect to $V_{rms}$, i.e., $dG/dV_{rms}$ (see main text and SM for more detail). The white dashed line represents a fit to $\gamma^{-1} V_{dip, I}(B)$, where $V_{dip,I}(B)$ is extracted from Fig. 2b, and the AC coupling factor $\gamma$ is $\approx$ 0.2. 
        }
        \end{figure}


Finally, we address heating effects stemming from the application of microwave radiation. 
To this end, we have studied the response of device A in the presence of high-frequency AC signals applied to an antenna located a few millimeters away from the nanowire. By sweeping the frequency of the signal at a given nominal power, we detect resonances at which a stronger effect on the dips is observed (see Supplemental Material). In Fig.~\ref{Fig4}a, we plot the impact of the AC signal amplitude, $V_{rms}$, at one of such resonances ($f = 3.492$ GHz) on the superconductor-to-normal transitions of the islands and leads. Interestingly, we observe that all three dips move to lower voltages with increasing $V_{rms}$, signaling that a progressively lower DC Joule power is required for turning the superconductors normal, due to heating from the AC signal. 
To model the observed behavior we assume that the microwave is absorbed by circuit elements converting it into an effective AC voltage source, yielding a total bias, $V_{tot} = V + V_{AC}\sin(2\pi f t)$. For frequencies much faster than the thermalization rate, $1/\tau_{th}$ a net Joule power, nominally given by $V^2/R+V_{AC}^2/2R$ when neglecting heating from excess current, is dissipated by the AC signal even though its voltage averages to zero over time. We simulate this using Eqs.~(\ref{eq:TotCurr}-\ref{eq:LeadPow}), assuming cooling by quasiparticle diffusion in leads and by phonons in the island with parameters from DC fittings, and using a smoothed step function to simulate the onset of excess current, and lastly, by using the average power of an AC cycle to self-consistently solve for the temperatures, $T_L$, $T_R$, and $T_I$. The results of these calculations are shown in Fig.~\ref{Fig4}b. The simulation captures very well the main qualitative features of the experimental data, including a splitting of the enhanced conductance at low-bias due to the onset of excess current, supporting that $f \gg 1/\tau_{th}$.In Fig.~\ref{Fig4}c, we evaluate the effect of the AC signal frequency on the low-bias dip. We notice a subtle yet important difference, namely the splitting of the finite bias $dI/dV$ dip for increasing $V_{rms}$ in the lower frequency ($f = 1$ MHz) measurement. Indeed, for frequencies slower or comparable to thermalization, we would expect $T_I$ to oscillate with the AC signal as the system seeks thermal equilibrium at each point of the AC cycle. In the limit of $f\ll 1/\tau_{th}$ this would amount to a convolution of the DC result with $F(V) = 1/\text{Re}\left(\sqrt{V^2-V_{AC}^2}\right)$, resulting in a dip splitting of $V = V_{dip,I} \pm \abs{V_{AC}}$ consistent with our observations. 

Assuming a harmonic solution to the heat-balance equation we obtain the following expression for the thermal relaxation time \cite{O'Neil2012Apr, Viisanen2018Mar},
\begin{equation}
\tau_{th} = \frac{6}{\pi}\frac{n\Sigma}{k_B^2\nu_{Al}}T^{n-2},
\end{equation}
with $n$ being the power of Eq.~(\ref{eq:PhCooling}), fitted to $n=6$ in experiment, and $\nu_{Al} = 2.15\hspace{0.1cm}10^{47}J^{-1}m^{-3}$ the Fermi density of states of aluminium. For $T = T_c = 1.35$~K, appropriate for a dip, we find $\tau_{th} \approx 10$~ns, consistent with our measurement of fast and slow dynamics. For further details see the supplemental material.


For the measurement taken at higher $f$, a critical value of the AC signal amplitude, $V_{rms}^{c}$, can be identified for which the low-bias dip disappears after merging at $V = 0$. This critical amplitude represents the full suppression of the superconductivity of the island driven solely by AC heating. To study this in more detail, we take measurements of the differential conductance at zero-bias, $G = dI/dV(V = 0)$, as a function of $V_{rms}$ and $B$. At $V_{rms}^{c}$, Josephson and Andreev processes are suppressed, leading to a drop in $G$.  This critical point is better visualized by plotting $dG/dV_{rms}(V_{rms})$, as shown in Fig.~\ref{Fig4}d, since this quantity approaches zero following the transition of the island. Strikingly, we observe that the $B$-field dependence of $V_{rms}^{c}$ is qualitatively identical to that of the DC voltage dip. This is underscored by the fit of the experimental data to $\gamma^{-1} V_{dip,I}(B)$, shown as dashed line, which is carried out by employing $V_{dip,I}(B)$ from Fig.~\ref{Fig2}b and yields an AC coupling factor, $\gamma \approx$ $0.2$, which would be $0.5$ for a perfect AC voltage source in the $f \gg 1/\tau_{th}$ regime. We therefore demonstrate that high-frequency AC signals give rise to Joule heating in a similar manner as in the DC case. 

We now briefly comment on neglected heating and cooling terms in our analysis. First, electron-phonon coupling is also present in the leads, but is estimated to be $\sim$ 100 times less efficient than quasiparticle diffusion and is therefore negligible. Also, the temperature gradients across the junctions, given by $T_{L/R}-T_{I}$ contribute to the cooling of the island and heating of the leads, as thermally activated quasiparticles cross the junction. This contribution is insubstantial for the leads, but can contribute to a cooling of about $10$\% of the phonon cooling for the island at $T_I=T_{c,I}$. For simplicity, we chose not to include this term, as its addition does not affect the quality of fits, and can therefore not be separated in experiment from phonon cooling, and would simply contribute to a slight rescaling of measured $\Sigma$. For more details on this see supplemental material. Future experiments will be directed at addressing these more subtle effects.



To conclude, we have performed a detailed study of heating effects in hybrid superconductor-semiconductor nanodevices by Joule spectroscopy. 
Our measurements show that the temperature of superconductors in a hybrid device can increase significantly with the Joule power deposited from DC electrical currents and/or microwave signals. Importantly, the temperatures of grounded and floating superconductors can differ greatly, being determined by their corresponding heat bottleneck, namely quasiparticle diffusion and electron-phonon coupling. We note that, as a direct consequence, mesoscopic superconducting islands display a much stronger susceptibility towards heating. 
Altogether, our observations suggest that non-equilibrium conditions and/or the presence of AC signals (including noise) can substantially impact the response of hybrid superconductor-semiconductor devices, especially in geometries with floating superconductors, by leading to increased local temperatures that can greatly exceed $T_{bath}$. While these effects have been rather overlooked in relation to hybrid devices, they can have implications for e.g., in geometries proposed for the realization of topological superconducting phases and related qubits \cite{aasen_milestones_2016, whiticar_coherent_2020, albrecht_exponential_2016, sherman_normal_2017, Valentini2022Dec, Sau2012Jul}. We note that while our experiment focuses on extreme conditions at which the superconducting elements of a device transition to the normal state, heating effects are in fact even more pronounced for lower (AC/DC) applied powers as both quasiparticle diffusion and electron-phonon coupling become less efficient with decreasing electron temperature.

\bigskip
\bigskip

When finalizing the preparation of this manuscript, we became aware of this related work \cite{kraphai_island_2023}.


\clearpage


\bigskip
\textbf{\large{Methods}}

\textbf{Sample fabrication and measured samples:} The devices discussed in the main text are based on full-shell InAs-Al nanowires and incorporate a mesoscopic superconducting island. By fitting the Little-Parks oscillations of the devices (see SM for more details), 
we estimate the inner diameter of the core InAs nanowire to be $\approx$ 150 nm ($\approx$ 110 nm) for device A (B), and the thickness of the Al shell to be $\approx$ 8 nm. 

The nanowires were deterministically transferred using a micro-manipulator from the growth chip to silicon substrates for further fabrication. Specifically, intrinsic (degenerately-doped) Si substrates with 300 nm-thick SiO$_2$ dielectrics were employed for device A (B).

Standard e-beam lithography (EBL) was used to define masks for wet etching segments of the epitaxial shell. Oxygen plasma descumming at 100 W for 75 s was carried out before immersing the samples in AZ326 MIF developer (containing 2.38\% tetramethylammonium hydroxide, TMAH) for 75 s at room temperature. For device A, two segments of $\approx$ 200 nm of the Al shell were etched, defining the superconducting island and two superconducting leads. By contrast, for device B, we removed completely the Al shell at the ends of the nanowire to form the island, and to allow the fabrication of normal metal leads at a subsequent process. Electrical contacts and side gates were fabricated by standard EBL techniques, followed by metallization by e-beam evaporation at pressures of $\sim 10^{-8}$ mbar. Ion milling is carried out prior to the evaporation to remove the native oxide of the Al shell (InAs nanowire) for device A (B). The evaporated electrical contacts consisted of superconducting Ti (2.5 nm)/Al (240 nm) for device A, and normal Cr (2.5 nm)/Au (180 nm) for device B. Note that, for device B, the Cr/Au leads are fabricated $\approx$ 200 nm away from the superconducting island, and that the doped Si substrate was used as a global back gate.

The main features discussed in this work have been observed in at least 4 devices similar to device A, and 2 similar to device B. 

\bigskip

\textbf{Measurements}: Our experiments were carried out in two different cryogenic systems: a dilution refrigerator with a base temperature of 20 mK, which was employed for measuring device A, and a $^3$He insert with a base temperature of 250 mK, employed in the measurement of device B. A description of the cryogenic filtering in these setups can be found in the Supplemental Material. The cryostat temperatures were measured by ruthenium oxide thermometers attached to the $^3$He pot and the mixing chamber of the above systems. 

All data related to device A was obtained by performing two-terminal voltage-bias transport measurements using standard lock-in techniques, whereby a voltage, $V$, is applied with a small low-frequency ($f = 117$ Hz) AC excitation, $dV$, and the current, $I$, and differential conductance, $dI/dV$ are measured. We have employed different amplitudes of the lock-in excitation depending on whether the measurement targeted resolving the low-bias dip only ($dV =$ 5 $\mu$V), or also the high-bias dips ($dV =$ 100-200 $\mu$V), as the contrast of the latter is much weaker. (Note: the $dV$ values listed above are the nominal ones, i.e., before subtracting the voltage drop on the cryogenic filters, as we explain below).

For device B, we have experimented with both two-terminal voltage-bias and current-bias measurements (see SM for further details). We have observed that the experimental data obtained with these two different schemes are completely equivalent, e.g., when comparing measurements plotted: (i) as a function of the applied voltage-bias or of the voltage drop across the device as a result of an applied current-bias, or (ii) as a function of the current resulting from an applied voltage-bias or of the applied current-bias. In the main text, we have opted to display the two-terminal current-bias measurements, owing to a reduced electrical noise in this configuration for our $^3$He insert setup, which translated into sharper $dI/dV$ dips. For this reason, $V$ in Fig. 3 refers to the total voltage drop across device B resulting from an applied current-bias, $I$. These measurements were taken with a lock-in excitation equal to $dI = 1$ nA at $f = 117$ Hz.

Microwave radiation from a signal generator (Rohde \& Schwarz SMW200A) was applied to an on-chip antenna located a few millimeters from device A. The coupling between the devices and the impedance of the microwave antenna is not precisely characterized. Therefore, the given $V_{rms}$ values in the main text are nominal values obtained as $V_{rms} \propto 10^{P/20}\sqrt{P_0}$ with $P_0 = 1\text{mW}$ and $P$ the nominal power in dBm units.
\bigskip

\textbf{Data processing.} The voltage drop on the total series resistance of our experimental setups, which is primarily due to cryogenic filters (2.5 k$\Omega$ per experimental line), has been subtracted for plotting the data shown in Figs. 1-4. 







\bigskip
\bibliography{bibliography}


\bigskip
\textbf{\large{Acknowledgments}}

We acknowledge funding by EU through the European Research Council (ERC) Starting Grant agreement 716559 (TOPOQDot), the FET-Open contract AndQC, by the Danish National Research Foundation (DNRF 101), Innovation Fund Denmark, the Carlsberg Foundation, and by the Spanish AEI through Grants No.~PID2020-117671GB-I00 and TED2021-130292B-C41, through the ``Mar\'{\i}a de Maeztu'' Programme for Units of Excellence in R\&D (CEX2018-000805-M) and the ''Ram\'{o}n y Cajal'' programme grant RYC-2015-17973.

\bigskip
\textbf{\large{Author contributions}}

A.I. fabricated the devices. A.I., I.C., M.G., and E.J.H.L. performed the measurements. G.O.S. and A.L.Y. developed the theory. G.O.S. performed the theoretical calculations. T.K. and J.N. developed the nanowires. All authors discussed the results. A.I., G.O.S., I.C., M.G., A.L.Y., and E.J.H.L, analyzed the experimental data and wrote the manuscript with input from all authors. E.J.H.L. proposed and guided the experiment.



\end{document}